\documentclass[%
 reprint,
superscriptaddress,
 amsmath,amssymb,
 aps,
]{revtex4-2}

\usepackage{graphicx}
\usepackage{dcolumn}
\usepackage{bm}

\usepackage[dvipsnames]{xcolor}
\usepackage{stackrel}
\usepackage{sidecap} 

\usepackage{amsfonts}

\usepackage[normalem]{ulem}

\usepackage{tabularx}

\newcommand{\kT}{k_BT}
\newcommand{\Dbulk}{D_0}
\newcommand{\VV}{\mathcal{U}}
\newcommand{\Vbox}{\mathcal{V}}

\newcommand{\Lbox}{L}
\newcommand{\dd}{\mathrm{d}}
\newcommand{\Dstar}{D_c}
\newcommand{\DN}{\Delta N}
\newcommand{\DNtt}{\brak{\DN^2(t)}}
\newcommand{\brak}[1]{\langle #1 \rangle}
\newcommand{\Nmean}{\brak{N}}
\newcommand{\Dhydro}{D_0^{\rm hydro}}
\newcommand{\Nvar}{\brak{N^2}-\Nmean^2}

\global\long\def\M#1{\boldsymbol{#1}}
\global\long\def\sM#1{\M{\mathcal{#1}}}
\global\long\def\Mob{\sM M}

\begin{document}

\preprint{APS/123-QED}

\title{The Countoscope: \\ Self and Collective Diffusion Coefficients by Counting Particles in Boxes}
\title{The Countoscope: \\ Self and Collective Dynamics by Counting Particles in Boxes}
\title{The Countoscope: \\ Self and Collective Diffusive Dynamics by Counting Particles in Boxes}
\title{The Countoscope: \\ Self and Collective Diffusive Dynamics without Trajectories}
\title{The Countoscope: \\ Self and Collective Dynamics without Trajectories}
\title{The Countoscope: \\ Measuring Self and Collective Diffusion without Trajectories}
\title{The Countoscope: \\ Measuring Self and Collective Dynamics without Trajectories}


\author{Eleanor K. R. Mackay}
\affiliation{Physical and Theoretical Chemistry Laboratory, South Parks Rd, Oxford, OX1 3QZ UK}

\author{Sophie Marbach}
\email{sophie.marbach@cnrs.fr}
\affiliation{CNRS, Sorbonne Universit\'{e}, Physicochimie des Electrolytes et Nanosyst\`{e}mes Interfaciaux, F-75005 Paris, France}
\affiliation{Courant Institute of Mathematical Sciences, New York University, New York, NY, USA}
\thanks{These three authors contributed equally}

\author{Brennan Sprinkle}
\email{bsprinkl@mines.edu}
\affiliation{Applied Math and Statistics, Colorado School of Mines, 1500 Illinois St, Golden, CO 80401}
\thanks{These three authors contributed equally}

\author{Alice  L. Thorneywork}
\email{alice.thorneywork@chem.ox.ac.uk}
\affiliation{Physical and Theoretical Chemistry Laboratory, South Parks Rd, Oxford, OX1 3QZ UK}
\affiliation{Cavendish Laboratory, Department of Physics, University of Cambridge, JJ Thomson Avenue, Cambridge CB3 0HE, UK}
\thanks{These three authors contributed equally}

\date{\today}

\begin{abstract}
Driven by physical questions pertaining to quantifying particle dynamics, microscopy can now resolve complex systems at the single particle level, from cellular organisms to individual ions. Yet, available analysis techniques face challenges reconstructing trajectories in dense and heterogeneous systems where accurately labelling particles is difficult. Furthermore, the inescapable finite field of view of experiments hinders the measurement of collective effects. 
Inspired by Smoluchowski, we introduce a broadly applicable analysis technique that probes dynamics of interacting particle suspensions based on a remarkably simple principle: counting particles in finite observation boxes. 
Using colloidal experiments, advanced simulations and theory, we first demonstrate that statistical properties of fluctuating counts can be used to determine self-diffusion coefficients, so alleviating the hurdles associated with trajectory reconstruction.
We also provide a recipe for practically extracting the diffusion coefficient from experimental data at variable particle densities, which is sensitive to steric and hydrodynamic interactions. 
Remarkably, by increasing the observation box size, counting naturally enables the study of collective dynamics in dense suspensions. Using our novel analysis of particle counts, we uncover a surprising enhancement of collective behaviour, as well as a new length scale associated with hyperuniform-like structure. Our counting framework, the Countoscope, thus enables efficient measurements of self and collective dynamics in dense suspensions and 
opens the way to quantifying dynamics and identifying novel physical mechanisms in diverse complex systems where single particles can be resolved.


\end{abstract}

\maketitle

A century ago, Smoluchowski proposed that diffusion coefficients in particle suspensions could be measured by probing correlations in the number of particles within the observation field of a microscope -- a number that fluctuates with time as particles stochastically jump in and out of the field due to Brownian motion~\cite{islam2004einstein,smoluchowski1916studien,chandrasekhar1943stochastic,svedberg1911neue,smoluchowski1916drei}. The simplicity of this idea made it experimentally accessible, and within a few years, Smoluchowski's theory was validated experimentally for dilute systems~\cite{westgren1918koagulation,westgren1916bestimmungen}.

\begin{figure*}[htp]
\includegraphics[width=0.9\linewidth]{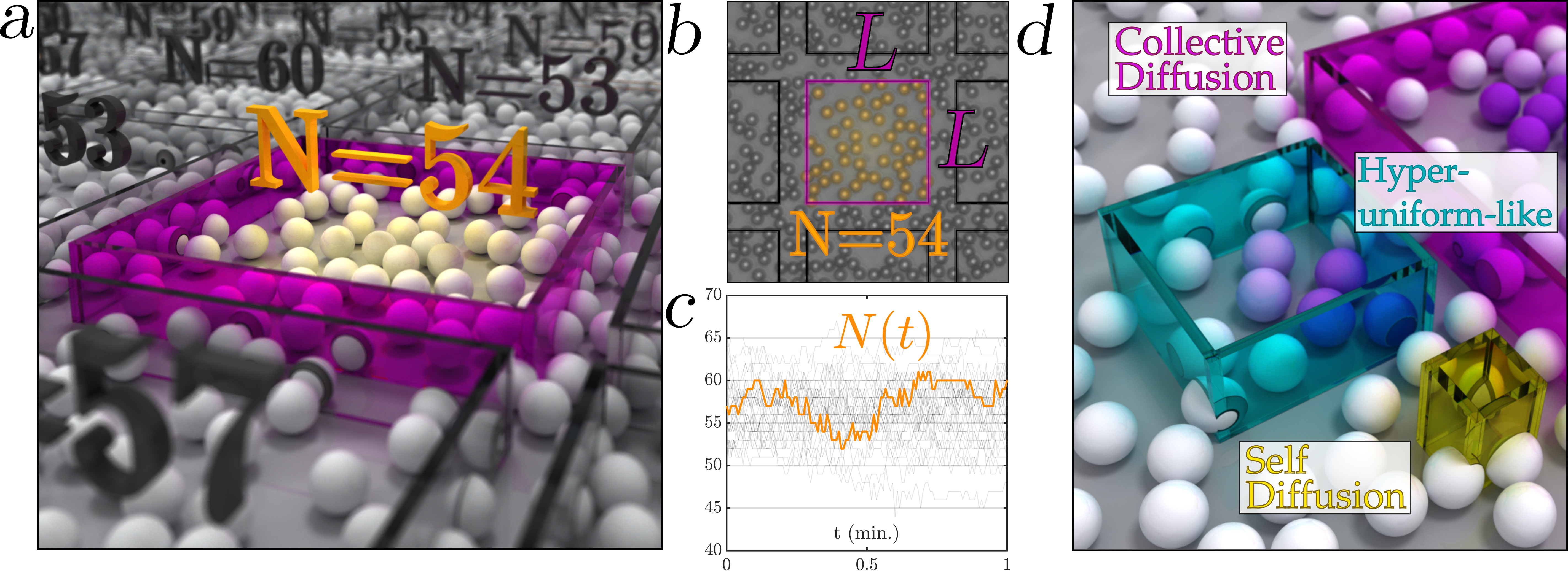}
\caption{\textbf{The Countoscope to probe particle dynamics}: (a) Numerical setup with freely diffusing colloids sedimented on a plane and pink observation box. (b) Image of the quasi-2D colloidal experiment at $\phi = 0.34$ with superimposed boxes. (c) Example of particle number fluctuations observed in a box of size $L = 32~ \mu$m at $\phi = 0.34$, from simulations. Counting particles does not require positions to be linked into trajectories, only that the position of each particle is sufficiently well resolved with respect to the box edges. (d) Boxes with different sizes $L$ are associated with different phenomena in systems at high packing fractions. }
\label{fig:fig1}
\end{figure*}

Since then, widespread interest 
in particulate suspensions has motivated the development of many techniques to measure diffusion coefficients~\cite{jameson2009fluorescence}. Broadly, these can be split into two categories: microscopy, in which individual particle positions can be resolved; and spectroscopy-like techniques, such as Dynamic Light Scattering~\cite{berne2000dynamic}, Fluorescence Correlation Spectroscopy~\cite{elson1974fluorescence} and Differential Dynamic Microscopy~\cite{cerbino2008differential}, where resolution is usually insufficient to identify individual particles. 
Spectroscopy implicitly relies on Smoluchowski's idea 
by correlating the light intensity received within the field of view, or at different length scales in Fourier space, as a proxy for number fluctuations. Yet, these methods require detailed models to unambiguously map intensity to particle properties, which are hard to define in dense systems with multiple-scattering~\cite{hassan2015making,rose2020particle}.


Microscopy gives direct access to particle configurations but, unlike spectroscopy, typically does not exploit density fluctuations to probe dynamics.
Instead, particle positions are found, and a linking algorithm is used to connect positions into trajectories, $x(t)$~\cite{crocker1996methods}. 
As proposed by the seminal works of Einstein and Perrin~\cite{einstein1905molekularkinetischen,perrin2014atomes}, the diffusion coefficient, $\Dbulk$, is then inferred by computing the particle's mean-square displacement from the trajectories, since $\brak{\Delta x^2(t)} = \langle (x(t)-x(0))^2 \rangle = 2\Dbulk t$. As such, 
the concept of explicitly counting particles as an experimental analysis tool, as introduced by Smoluchowski, is largely forgotten. The power of this approach for analysis in modern microscopy settings is thus unexplored. 

Our ability to resolve single-particle behaviour in complex systems has improved dramatically in recent 
years, with super-resolution and interferometric scattering microscopy now resolving individual molecular dynamics~\cite{comtet2021anomalous,ronceray2022liquid}.
In complex cases such as these, while particle positions can be resolved, linking algorithms face challenges. For instance, linking algorithms cannot cope with particles appearing and disappearing due to their diffusion outside of the field of view, bleaching, cellular division, or substantial drift~\cite{manzo2015review,ghosh2022cross,ollion2023distnet2d}. In addition, linking becomes ambiguous in dense or heterogeneous systems, as it relies on distance cutoffs~\cite{manzo2015review,deforet2012automated,rose2020particle}. 
Single-particle microscopy is, therefore, poised for new analysis techniques to probe dynamics~\cite{miles2023inferring,gnesotto2020learning,ollion2023distnet2d}.

Here, we introduce a technique, the \textit{Countoscope}, which exploits fluctuating particle counts in finite observation volumes to infer dynamic properties for interacting particle suspensions (Fig.~\ref{fig:fig1}). We use varying observation box sizes, going beyond Smoluchowski's initial idea, to explore different dynamical effects in many-body interacting systems (Fig.~\ref{fig:fig1}d). 
 With a combination of bright-field experiments on quasi-2D colloidal suspensions, simulations, and theory, we first demonstrate that fluctuating counts 
 can be used to determine self-diffusion coefficients while removing the challenges associated with trajectory reconstruction (Sec. I). We provide a recipe for practically extracting the diffusion coefficient from suspensions at variable particle densities, that is sensitive to steric and hydrodynamic interactions. Second, we uncover novel collective behavior of our deceptively simple 2D colloidal suspension, by increasing the observation box size (Sec. II). A new length scale that characterizes a transition between hyperuniform-like and collective states emerges from our novel analysis of counts. We also find a surprising enhancement, up to an order of magnitude, of collective dynamics, that we can attribute to hydrodynamic interactions. 
 Overall, we demonstrate the potential to infer dynamic properties from particle counting; a technique with clear extensions to more complex systems where single-particle microscope images are possible, including motile particles, biological or artificial, in 2D as well as in 3D. 




\section{Self-diffusion coefficients without trajectories}

\subsection{Particle number fluctuations in boxes at low densities}

Our experimental hard sphere model system consists of monolayers of colloidal particles  (diameter $\sigma= 2.8~\mu m$) gravitationally confined to the base of a glass cell ~\cite{thorneywork2014communication,Thorneywork2015}. 
Samples are imaged using a custom-built inverted microscope 
%
with particle positions acquired from images using standard particle tracking protocols~\cite{crocker1996methods,allan_daniel_b_2021_4682814}. The full range of packing fractions associated with the fluid phase, $\phi = 0.02 - 0.66$, is explored~\cite{Thorneywork2017}.   
We simulate colloidal suspensions using the  Brownian dynamics method described in Ref.~\citenum{sprinkle2020driven} that includes hydrodynamic lubrication. All simulation parameters, such as particle diameter and temperature, are set to their experimentally measured values, and steric forces are modelled using the firm potential in Eq.~(31) of Ref.~\citenum{sprinkle2020driven}. Hydrodynamic interactions can be turned off in our simulations by setting the hydrodynamic mobility matrix to the identity matrix. 

\begin{SCfigure*}
\centering
\includegraphics[width=1.3\linewidth]{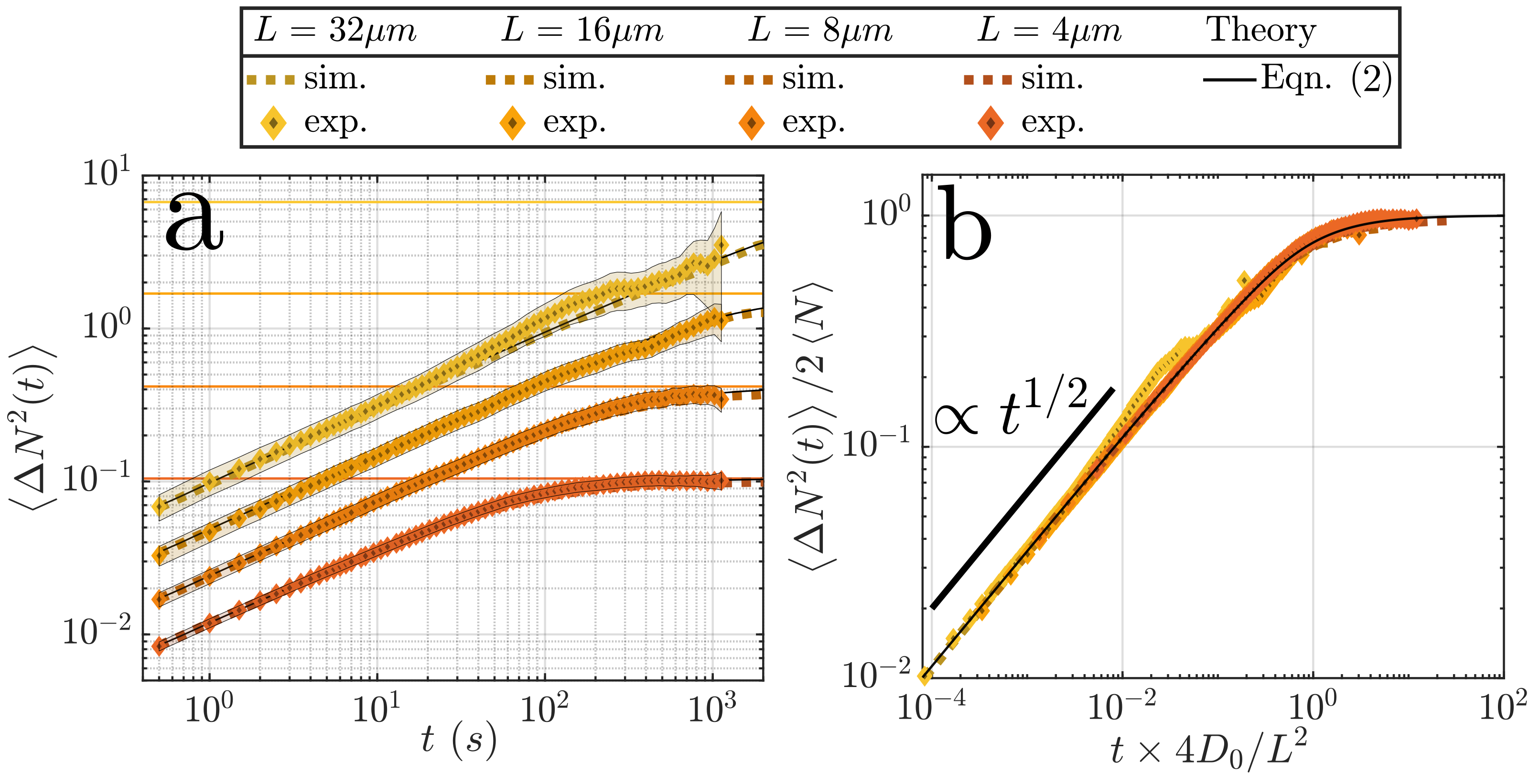} \hspace{1mm}
\caption{\textbf{Particle number fluctuations in the dilute regime.} (a) The mean squared change in particle number at $\phi=0.02$ for experiments (points) and simulations (dashed line). Colours indicate increasing box sizes going from orange to yellow and solid coloured lines correspond to $2\Nmean$ for each box size. (b) Data from (a) are rescaled in $y$ by the average number of particles in a box, $\Nmean$, and in time by the diffusive time scale $\Lbox^2/\Dbulk$. The solid black line shows the prediction of Eq.~\eqref{eq:fluct1}. Error bars on experimental data are shown as transparent regions and represent 95\% confidence intervals due to statistical error on the averaging over multiple boxes and time origins, which is the dominant source of error. }
\label{fig:fig2}
\end{SCfigure*}

For both experiments and simulations, we characterise fluctuations 
by counting the number of particles within boxes of size $\Lbox\times \Lbox$ over time. 
This number, $N(t)$, fluctuates between discrete values as a consequence of particles diffusively entering and exiting the box (Fig.~\ref{fig:fig1}c). 
For each box we compute the mean squared change in particle number 
\begin{equation}
\begin{split}
\label{eq:nMSD}
    \DNtt  &= \brak{\left(N(t) - N(0)\right)^2} \\
    &= 2 (\brak{N^2} - \Nmean^2) - 2 (\brak{N(t) N(0)} - \Nmean^2) \\
   &=  2 (\brak{N^2} - \Nmean^2) - 2 C_N(t),
\end{split}
\end{equation}
where $C_N(t) = \brak{N(t) N(0)} - \Nmean^2$ is the time correlation function for the particle number and $\langle \cdot \rangle$ indicates an average over all boxes and time origins within the acquisition. 
Since $\DNtt$ calculates \textit{differences} between frames, it yields more accurate results than the correlation function $C_N(t)$ alone, \textit{e.g.} by automatically removing stuck particles. 

In Fig.~\ref{fig:fig2}-a, we plot the mean squared change in particle number, $\DNtt$, for different box sizes in the dilute regime ($\phi = 0.02$). 
Phenomenologically, $\DNtt$, increases with time and eventually plateaus, with experiments (diamonds) and simulations (dashed lines) in perfect agreement within error bars.
Initially, number fluctuations increase due to particles entering or exiting the box. Over time this eventually results in complete exchange of particles inside the box with those outside and we observe a plateau. The number of particles at long times is therefore uncorrelated with that in the initial configuration, \textit{i.e.} $C_N(t) \simeq 0$
and, so, from Eq.~\eqref{eq:nMSD}, the plateau corresponds to the variance $\DNtt \simeq 2 (\brak{N^2} - \Nmean^2)$. For low packing fractions, we expect the variance to be equal to the average particle number $(\brak{N^2} - \Nmean^2) = \Nmean$~\cite{hansen2013theory}, which suggests a rescaling of $\DNtt$ by $2\Nmean$. The time required to reach this plateau depends on the characteristic time to exchange particles. For diffusing particles, this is simply the time to diffuse in and out of the box, $\Lbox^2/4\Dbulk$. Rescaling particle number fluctuations and time accordingly, we obtain a remarkable data collapse (Fig.~\ref{fig:fig2}-b). 


We uncover the scaling function accounting for this behaviour using stochastic density field theory (sDFT)~\cite{dean1996langevin,kawasaki1998microscopic}, which models the fluctuating evolution of the particle number density.
Assuming relative density fluctuations are small, we
obtain 
(Supp. Mat. Sec.~2.1)
\begin{equation}
\begin{split}
    &\langle \DN^2(t) \rangle = 2 \Nmean \left( 1 - \left[ f\left( \frac{4 \Dbulk t}{\Lbox^2}\right) \right]^2 \right), \,\,\,\,\mathrm{where}\\
    &  \,\, f\left(\tau = \frac{4 \Dbulk t}{\Lbox^2}\right) = \sqrt{\frac{\tau}{\pi}} \left( e^{-1/\tau} - 1\right) + \mathrm{erf} \left( \sqrt{1/\tau} \right).
    \end{split}
    \label{eq:fluct1}
\end{equation}
Eq.~\eqref{eq:fluct1} 
agrees remarkably well with our experimental and simulation data (Fig.~\ref{fig:fig2}b). 
The formalism can be used similarly in $d = 1,2,3$ dimensions, where the square factor in Eq.~\eqref{eq:fluct1} becomes $\left[ f(4D_0 t/L^2) \right]^d$. This allows one to recover predictions such as Smoluchowski's on non-interaction Brownian suspensions in 1D ~\cite{smoluchowski1916studien}, or on ionic suspensions in 3D~\cite{minh2023ionic}.
Curiously, the initial increase in time of $\DNtt$ is subdiffusive, growing as $\sqrt{t}$, despite particle motion in the dilute regime being strictly diffusive. As previously predicted in 1D for non-interacting particles~\cite{marbach2021intrinsic,dean2023effusion,di2023current}, this subdiffusive scaling emerges from particle crossings in and out of a box,  
and, therefore, does not depend on the system's dimension. 

Expanding Eq.~\eqref{eq:fluct1} at early times predicts
\begin{equation}
\langle \DN^2(t) \rangle = \langle \left(N(t)-N(0)\right)^2\rangle = \frac{8}{\sqrt{\pi}} \Nmean \sqrt{\frac{\Dbulk t}{\Lbox^2}}.
    \label{eq:fluctaround0}
\end{equation}
Eq.~\eqref{eq:fluctaround0} is the analogue of the mean squared displacement~\cite{einstein1905molekularkinetischen} $\brak{\Delta x^2(t)} = 2 \Dbulk t$, but here for particle number fluctuations. Eq.~\eqref{eq:fluctaround0} can thus be used to fit the particle number fluctuations at early times to determine the self-diffusion coefficient $\Dbulk$ simply by counting, and we come back to further explanations on the fitting procedure in Sec. I-C. Crucially, our method accurately reproduces $\Dbulk$ obtained from a conventional mean squared displacement within error bars (Table~\ref{tab:fits}, $\phi = 0.02$). Compared to the Stokes-Einstein prediction for $D_0$ in the bulk, our measured diffusion coefficient is reduced by about 70\% due to increased hydrodynamic friction with the base glass cell~\cite{sprinkle2020driven}.

\begin{SCfigure*}
\centering
\includegraphics[width=1.3\linewidth]{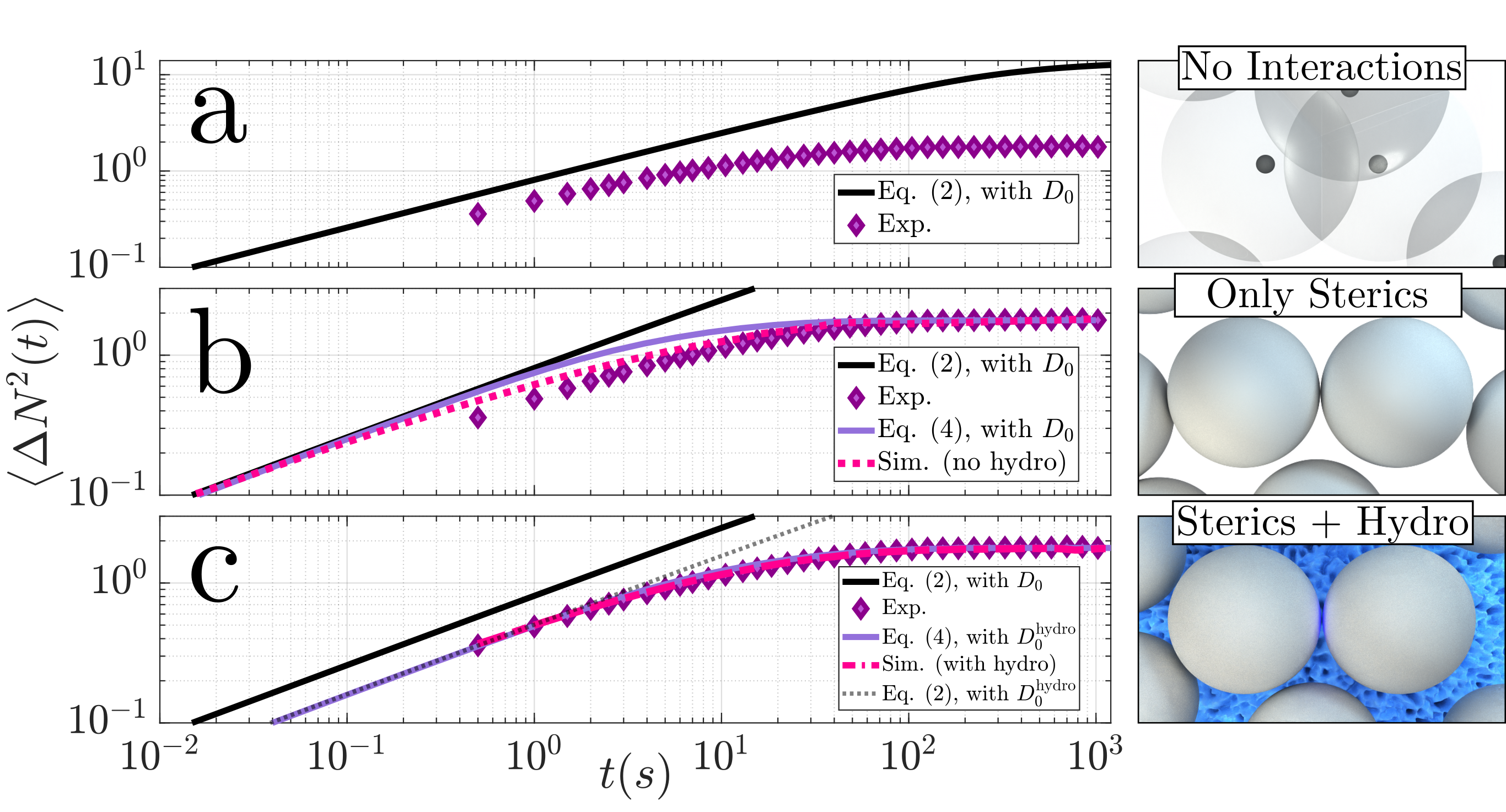} \hspace{1mm}
\caption{\textbf{Sensing steric and hydrodynamics effects} The mean square change in particle number for a box size $L = 8~\mathrm{\mu m}$ and packing fraction $\phi = 0.66$. Panels compare experimental data (diamonds) to results from different simulation and modeling scenarios (lines) in which: (a) particles do not interact (b) particles only have steric interactions (c) particles feel both steric and hydrodynamic interactions. For comparison, theory without interactions (black line) is presented in each panel, and corresponds to Eq.~\eqref{eq:fluct1}. In (b), the slight deviation between simulations without hydrodynamics and Eq.~\eqref{eq:fluctaround0} is likely due to the assumption of small relative density fluctuations in the theory. }
\label{fig:fig3}
\end{SCfigure*}

\subsection{Sensing steric and hydrodynamic interactions at intermediate densities}

Few experiments can be approximated as a non-interacting system.
We, therefore, move beyond Smoluchowski's work, and our own theoretical investigations in Refs.~\cite{minh2023ionic,marbach2021intrinsic}, by developing the counting methodology in the \textit{dense} regime. In Fig.~\ref{fig:fig3}a we show $\DNtt$ with $L=8\mu$m for experiments at $\phi = 0.66$, which is just below the freezing transition \cite{Thorneywork2017}. Here, in contrast to systems at low  $\phi$, experimental data significantly deviates from our analytic expression, Eq.~\eqref{eq:fluct1}. This clearly demonstrates that particle number fluctuations are sensing interparticle interactions. To account for this effect theoretically, however, we must understand which type of interactions, \textit{i.e.} hydrodynamic and/or steric, are at play.

To address this, we first include pair-wise steric interactions in the theory (Supp. Mat. Sec.~2.2). Conveniently, to implement this within sDFT, the detailed form of the pair-wise interactions need not be specified; the only ingredient required is the static structure factor $S(k)$, which is well established analytically for this colloidal system~\cite{thorneywork2018structure} (Eq.~\eqref{eq:Sk} in Methods). We thereby obtain an improved version of Eq.~\eqref{eq:fluct1} as
\begin{equation}
     \langle \DN^2(t) \rangle = 2\Nmean  \displaystyle \int \frac{k \dd k}{(2\pi)^2} f_{\Vbox}(k) S(k) \left( 1 - e^{-\frac{\Dbulk k^2 t}{S(k)}} \right),
        \label{eq:fluctk}
    \end{equation}
where the function $f_{\Vbox}(k)$ is a characteristic area
\begin{equation}
 f_{\Vbox}(k) = \Lbox^2 \int \dd \theta \left( \frac{ 2\sin \left( \frac{k \Lbox \cos \theta}{2} \right)}{k \Lbox \cos \theta}\right)^2 \left( \frac{ 2\sin \left( \frac{k \Lbox \sin \theta}{2} \right)}{k \Lbox \sin \theta}\right)^2.
    \label{eq:fluctkVk}
\end{equation}
Eq.~\eqref{eq:fluctk} simplifies to Eq.~\eqref{eq:fluct1}, when particle interactions can be neglected, $S(k) \equiv 1$. 
Eq.~\eqref{eq:fluctk} significantly improves the agreement with experiments (Fig.~\ref{fig:fig3}b) as the theory now reproduces the lower plateau value. Here, we also show the corresponding result from simulations with only steric repulsion between particles, which also reproduces the plateau correctly. However, the early-time behaviour in experiments is not well captured with steric interactions alone. 

In Fig.~\ref{fig:fig3}c, we show results for simulations that now also include hydrodynamic interparticle interactions and perfectly reproduce experimental data at all times. This demonstrates that particle number fluctuations can only be understood by accounting for both sterics and hydrodynamics. 
Including hydrodynamic interactions in the theory is non-trivial~\cite{bernard2023analytical}. However, an immediate improvement to Eq.~\eqref{eq:fluctk} can be achieved by replacing the infinite dilution diffusion coefficient, $\Dbulk$, with the short-time diffusion coefficient, $\Dhydro$, which is affected by hydrodynamic interactions~\cite{Thorneywork2015}.
More specifically, $\Dhydro<\Dbulk$ at high packing fractions,
since hydrodynamics add friction between nearby spheres~\cite{goldman1967slow,sprinkle2020driven}. With this correction, our model Eq.~\ref{eq:fluctk} reproduces experimental data accurately for all times (Fig.~\ref{fig:fig3}-c and Supp. Mat. Sec. 1.3). 

\subsection{Guidelines to measure the short-time self-diffusion coefficient from counts}

Establishing the effect of interparticle interactions on particle number fluctuations reveals that for dense systems we can again use particle number fluctuations to quantify dynamics. For interacting systems, at short times, expanding Eq.~\eqref{eq:fluctk} yields Eq.~\eqref{eq:fluctaround0}, where $\DNtt$ is governed by the short-time self-diffusion coefficient including interparticle hydrodynamics $\Dhydro$ (dotted line in Fig.~\ref{fig:fig3}-c). This suggests a universal procedure to extract $\Dhydro$ for interacting systems as
\begin{equation}
    D_0^{\text{hydro}}  = \frac{\pi L^2}{ \Delta t} \left(\frac{\langle  \Delta N^2(\Delta t) \rangle}{ 8\langle N \rangle}  \right)^2  
    \label{eq:fit}
\end{equation}
where $\Delta t  = 0.5~\mathrm{s}$ is the time step between images. The values of $\Dhydro$ obtained for various packing fractions are reported in Table~\ref{tab:fits} and are accurate, within error bars, in reproducing that obtained from mean squared displacements. This confirms that the counting method is indeed suited to measure individual dynamics in situations where trajectory-based analysis is limited. Here, the number fluctuations are evaluated only at the smallest time interval $\langle  \Delta N^2(\Delta t) \rangle$. This is consistent with approaches to accurately fit to the mean-squared displacement for systems with small localization uncertainty ~\cite{michalet2010mean}. Alternatively, one could use Eq.~\eqref{eq:fluctaround0} (or even Eq.~\eqref{eq:fluct1}) to fit data over multiple time points. 
Conveniently, the estimator Eq.~\eqref{eq:fit} does not require knowledge of the structure factor $S(k)$, as sterics only affect fluctuations at long times.  

We next outline some practical guidelines for measuring the short-time self-diffusion coefficient from fluctuating counts. 
Most importantly, $\Dhydro$ is determined by averaging Eq.~\eqref{eq:fit} over a range of box sizes, $L \in [L_{\rm min}, L_{\rm max}]$, and this range should be chosen carefully. 
$L_{\rm min}$ should be large enough to resolve the early time regime in the correlation function. $L_{\rm max}$ should not be too large as an image can be paved with fewer large boxes than small ones, resulting in increased statistical error on the correlation function for larger $L$. This imposes a hard minima on the time step $\Delta t$ which should be short enough to resolve the early time regime in the largest possible box, which is the size of the field of view $ L_{\rm field}$, \textit{i.e.} $\Delta t \ll L_{\rm field}^2/D_0$. In practice, provided this rather light experimental requirement is fulfilled, not all but a suitable range of box sizes should be accessible for any dataset. 
Using longer datasets with more time points continuously improves statistics by reducing the error on the correlation function. We find, however, that approximately $150$ time points (corresponding to $1.25~\mathrm{min}$ at 2~fps) are sufficient to get accurate results (Table S2). 
We provide more details on systematically choosing the box size range in the Methods.


On small boxes, errors on $D_0^{\rm hydro}$ are typically larger, as these boxes approach the plateau at shorter times, and so deviations from the $\sqrt{t}$ scaling occur sooner (see Fig.~S7). Significant improvement can be obtained by expanding Eq.~\eqref{eq:fluctk} at short times to second order. This yields an updated formula 
\begin{equation}
    D_0^{\text{hydro}}  = \frac{\pi L^2}{4 \Delta t} \left( 1 -\sqrt{1 - \frac{\langle  \Delta N^2(\Delta t) \rangle}{2 \langle N \rangle} } \right)^2 .
    \label{eq:fit2ndorder}
\end{equation}
to estimate $\Dhydro$. Eq.~\eqref{eq:fit2ndorder} converges to Eq.~\eqref{eq:fit} in the limit of early times, when $\langle  \Delta N^2(\Delta t) \rangle \ll \Nmean$.  This second order estimator allows one to average data over a wider range of box sizes and yields even more accurate results (Table~\ref{tab:fits}). 

Next, we use a sensitivity analysis to explore the extent to which both estimators 
are robust to common experimental artefacts. 
From this, we determine that our method is relatively insensitive to systematic drift $v$ in the data (Table S3), providing the time resolution is fast enough ($\Delta t \leq D_0^{\rm hydro}/v^2$). Lack of subpixel resolution in positions also does not negatively impact our estimates (Table S5). A small percentage of stuck particles directly affects the uncertainty on $\Dhydro$ by introducing uncertainty on $\Nmean$. A $0.5-1\%$ fraction of stuck particles -- as we estimate in our setup -- results in a $0.5-1\%$ error on $D_0^{\text{hydro}}$, however, which is much smaller than other sources of error. In contrast, a comparatively small fraction of $0.1\%$ blinking particles changes the shape of the correlation functions and thereby reduces the accuracy of our estimate of $\Dhydro$ (Table S4). This final effect arises from the sensitivity of counting to specific system dynamics, and in fact, opens up the possibility of future extensions of the method to quantify different dynamic phenomena simultaneously. 
Overall, the above analysis defines rather light constraints, making us hopeful that a diversity of systems and setups beyond bright field microscopy should be able to access diffusive properties with fluctuating counts.

\begin{table}[h!]
    \begin{tabular}{p{1.5cm}||p{1.5cm}|p{1.5cm}|p{1.5cm}}
        \textbf{Method} & $\phi=0.02$ & $\phi=0.34$ & $\phi=0.66$ \\
        \hline
        $\langle \Delta x^2(t) \rangle$ estimator & $4.34$ ($\pm0.13$) & $3.16$ ($\pm0.10$) & $1.85$ ($\pm0.06$) \\
        \hline
$\langle \DN^2(t) \rangle$ 2nd order estimator & $4.36$ ($\pm 0.13$)   & $3.16$ ($\pm0.10$)  & $1.83$ ($\pm 0.06$)  \\
\hline
$\langle \DN^2(t) \rangle$ 1st order estimator & $4.32$ ($\pm 0.13$)   & $3.05$ ($\pm 0.09$)  & $1.79$ ($\pm 0.05$) \\ 
    \end{tabular}
    \caption{Short-time self-diffusion coefficient $D_0^{\text{hydro}} (\times10^{-2}~\mathrm{\mu m^2/s}$) obtained through 3 different methods: (1) $D_0^{\text{hydro}} = \langle \Delta x^2(\Delta t) \rangle/2\Delta t$ using mean-square displacements (MSD) -- averaged over all particle trajectories and time origins within a trajectory; (2) $D_0^{\text{hydro}}$ using the Countoscope Eq.~\eqref{eq:fit2ndorder} and (3) Eq.~\eqref{eq:fit}. Both (2) and (3) are obtained through an average over a range of box sizes and time origins. See Methods for details. Error bars reflect the $3\%$ experimental uncertainty on converting a displacement from pixels to $\mu m$. The duration of the movies used for all analysis was $20~\mathrm{min}$ at 2fps.} 
    \label{tab:fits}
\end{table}

\begin{SCfigure*}
\centering
\includegraphics[width=1.35\linewidth]{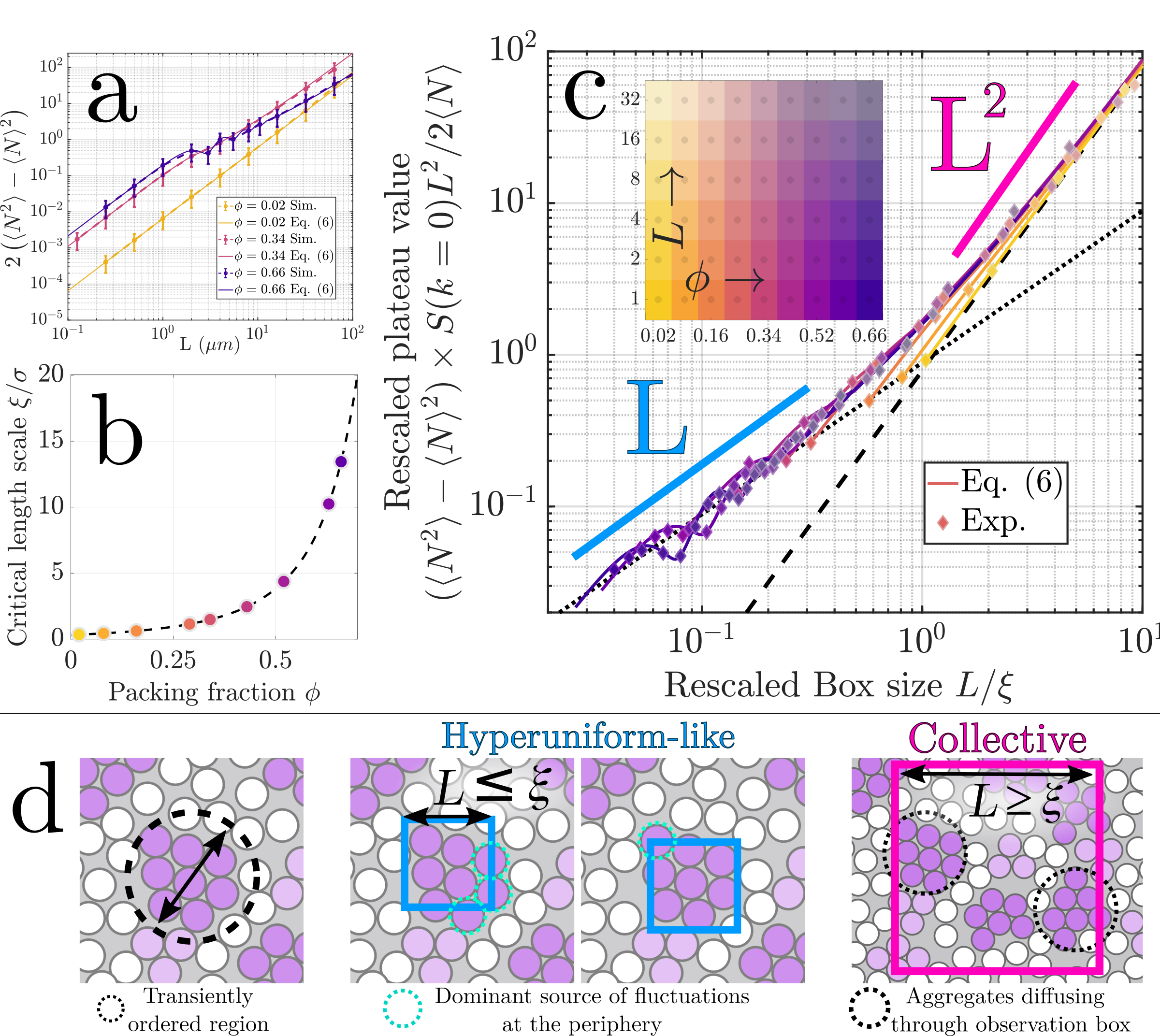} \hspace{1mm}
\caption{\textbf{Hyperuniform-like behaviour for small boxes defines a new length scale underlying colloidal fluctuations.} (a) Particle number variance $\Nvar$ with box size $\Lbox$, for increasing packing fractions $\phi$. Data is from simulations including hydrodynamics (points) with theory (lines) from Eq.~\eqref{eq:plateau}. Experimental data (not shown for clarity) also agree; (b) Crossover length scale $\xi$ (defined in Eq.~\eqref{eq:xi} with Eq.~\eqref{eq:sk0}) with $\phi$. Dots correspond to calculated values at the experimental packing fractions used in this study. (c) The rescaled variance shows a collapse of the data and the emergence of a hyperuniform-like regime for boxes smaller than $\xi$.  Experiments (diamonds) with theory (lines) from Eq.~\eqref{eq:plateau}. Boxes sizes are limited to $L \gtrsim \sigma$; when $L \ll \sigma$, we recover $\Nvar = \Nmean $ (Supp. Mat. Sec. 1.1). (d) Schematic illustrating the hyperuniform-like regime for intermediate box sizes. Dark (light) purple disks highlight large (small) groups of particles that are transiently in contact. }
\label{fig:Plateaus}
\end{SCfigure*}
\section{Fresh insights on collective properties}

We now turn to investigate collective dynamics at high packing fractions, which are especially challenging to resolve with trajectories.  To measure collective dynamics in dense states, we must first understand the static signatures of collective effects. These are most clearly observed in variations of the plateau value. 



\subsection{Particle interactions induce a hyperuniform-like regime in the number variance at small scales}

 The plateau value corresponds to twice the particle number variance, $\langle \Delta N^2(t\rightarrow \infty) \rangle = 2[\Nvar]$ and, by setting $t \rightarrow \infty$ in Eq.~\eqref{eq:fluctk}, we find
\begin{equation}
    \frac{\Nvar}{\Nmean} = \int \frac{k \dd k}{(2\pi)^2} f_{\Vbox}(k) S(k).
    \label{eq:plateau}
\end{equation}
Therefore, the variance over different box sizes, and consequently the plateau value, probes the structure factor at different scales. 

Fig.~\ref{fig:Plateaus}a shows the plateau value as a function of box size for three different values of $\phi$, demonstrating perfect agreement between simulation (with hydrodynamic interparticle interactions) and theory. 
Here, oscillations in the variance at high $\phi$ are a direct consequence of the increasingly well-defined short-range structural order in dense fluids; steadily increasing the box size naturally probes regions of high and low particle density arising from successive coordination shells~\cite{thorneywork2014communication}. 

Surprisingly, two regimes emerge from our theoretical model for the scaling of the variance with box size, for $\Lbox \ge \sigma$. When $L \gg \sigma$, a Taylor expansion of Eq.~\eqref{eq:fluctkVk} shows  
$\Nvar=S(k=0)\Nmean=S(k=0)(4\phi/\pi\sigma^2)L^2$ (Supp. Mat. Sec. 2.3). This recovers the well-established link between number variance and the compressibility $\chi_T$ of the system, as $\Nvar/\Nmean = S(k=0) = \rho k_B T \chi_T$~\cite{gomer1990diffusion,hansen2013theory}, with $\rho$ the mean particle number density.  In contrast, when $L \sim \sigma$, fluctuations scale with the \textit{length} of the box and not the area, as
$\Nvar = \Nmean \sigma/\pi L  = (4 \phi/\pi^2 \sigma^2) \sigma L$.
The crossover between the two regimes  defines a unique length scale
\begin{equation}
    \xi =   \frac{1}{\pi} \frac{\sigma }{S(k=0)},
    \label{eq:xi}
\end{equation}
which we plot as a function of $\phi$ in Fig.~\ref{fig:Plateaus}b. Rescaling the number variance by $\langle N \rangle \times (\sigma \xi/\pi L^2)$ and box size by $\xi$ shows perfect collapse of all experimental data (Fig.~\ref{fig:Plateaus}c) and confirms the emergence of these two regimes.

This behaviour can be interpreted by considering the dominant contributions to number variance in a box. For a large box, number fluctuations are entropy-dominated, associated with random placement of \textit{particles} in space. Since the particles are hard spheres, they cannot interpenetrate, and configurations where particles would overlap are not accessible. At high packing fractions, this leads to transient regions with significant short range structural order (Fig.~\ref{fig:Plateaus}-d, pink box). The variance is, thus, reduced by a factor $S(k=0)$,
that accounts for the loss of conformations due to steric interactions, as $\Nvar \sim S(k=0)\Nmean = \rho S(k=0) L^2$. 

The surprising regime for small box sizes, where $\Nvar \sim L$, can instead be investigated within the framework of \textit{hyperuniformity}~\cite{hexner2017enhanced,torquato2018hyperuniform,martin1980charge,ghosh2017fluctuations,kim2005screening}. 
A 2D system is \textit{hyperuniform} if number fluctuations scale at most like $\Lbox$ in the limit of infinitely large boxes~\cite{torquato2018hyperuniform}. A perfect crystal of hard spheres is hyperuniform, however, hard-sphere liquids are not~\cite{torquato2018hyperuniform}. 
In spite of this, a hyperuniform-like scaling emerges for small boxes. Since this regime extends over broader ranges of box sizes for larger packing fractions, we suggest the scaling is a consequence of short-range structural order (Fig.~\ref{fig:Plateaus}-d, blue box, see also Fig.~S2). 
At length scales smaller than transiently ordered regions, number fluctuations 
are dominated by the placement of particles relative to the box boundary~\cite{wu2015search,ikeda2017large}. 
This results in a variance scaling as the box perimeter multiplied by the particle diameter $\Nvar~\sim~4L\, \sigma$. 

The characteristic box size where these number fluctuation regimes are equal defines the length scale $\xi$.
The presence of these two regimes is reminiscent of charged particles systems~\cite{van1979thermodynamics, martin1980charge, kim2005screening, lebowitz1983charge} or particles interacting with interfaces~\cite{patel2010fluctuations, rotenberg2011molecular, rego_understanding_2022}, however, we are unaware of previous accounts of the length scale, $\xi$, for hard sphere fluids \cite{hansen2013theory}. In particular, this length scale reaches up to$~13$ particle diameters in our densest system at $\phi = 0.66$, further 
than the decay of the pair correlation function $g(r)$ or the bond-orientational correlation function, $g_6(r)$, where $r$ is the interparticle distance (Supp. Mat. Sec. 1.2)~\cite{hansen2013theory}. 
The length scale, $\xi$, which governs the behaviour of fluctuations, is thus apparently unrelated to \textit{average} correlation length scales. We note, however, that the maximal extent of transiently ordered regions will be larger than the average value, suggesting that $\xi$ may instead reflect this maximal value.

\begin{figure*}[htp!]
\centering
\includegraphics[width=0.99\linewidth]{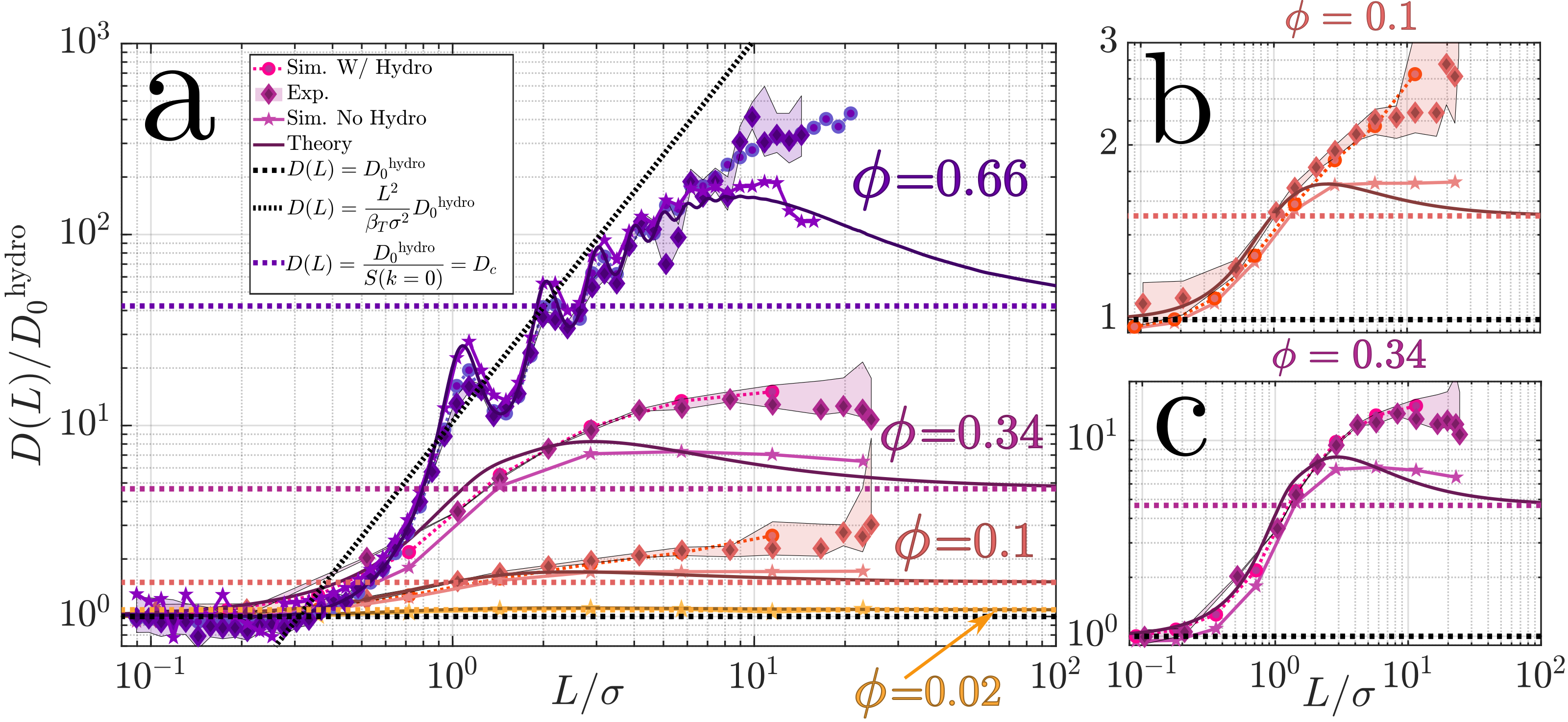}
\caption{\textbf{Box-size dependent diffusion coefficient measures individual and collective dynamics.} (a) Rescaled box-size dependent diffusion coefficient $D(L)/D_0^{\rm hydro}$ with rescaled box size $L/\sigma$, integrated from experimental and numerical data and theory via Eqs.~(\ref{eq:time}-\ref{eq:Dbox}). The rescaling $D_0^{\rm hydro} = D_0$ for numerical data without hydrodynamics and for theory. Different colours indicate varying packing fractions, going from low (yellow) to high (purple). (b-c) Same as (a) highlighting only (b) $\phi = 0.1$ and (c) $\phi = 0.34$ packing fractions.  Experimental data corresponds to 20h acquisition.
 Error bars, presented on experimental data only for simplicity, are propagated from errors in the timescale integrals -- see details in Supp. Mat. Sec. 1.4. }
\label{fig:Dstar}
\end{figure*}

\subsection{Direct measurements of collective diffusive dynamics}

Inspired by the different regimes unravelled for static properties at varying box sizes, we now explore dynamics beyond the short-time limit that is linked to self-diffusion. 
To this end, we define a timescale, $T(L)$,
to characterize the time when number fluctuations have relaxed -- visually, the time where $\DNtt$ reaches the plateau -- as 
\begin{equation}
\begin{split}
    T(\Lbox) &= 2 \int_0^{\infty}  \left(\frac{C_N(t)}{C_N(0)} \right)^2 \, \dd t \\
    &=  2 \int_0^{\infty}  \left(1 - \frac{1}{2}\frac{\DNtt}{\Nvar} \right)^2 \, \dd t. 
       \label{eq:time}
\end{split}
\end{equation}
Our definition means that for correlations that decay exponentially, $T$ would represent the timescale of the decay as $C_N(t)/C_N(0) = \exp(-t/T)$. 
Note that this definition of the relaxation timescale is different from Ref.~\cite{minh2023ionic}, where here a squaring factor in the integrand ensures that the integral converges, as the correlation function decays algebraically at long times, $C_N(t\rightarrow \infty) \sim 1/t$ (Supp. Mat. Sec. 2.5). 

In the absence of interparticle interactions, we find analytically from Eq.~\eqref{eq:fluct1} that $T$ corresponds to the time to diffuse across the box, $T(\Lbox) = \alpha_T \Lbox^2/4\Dbulk$ where $\alpha_T~=~2 \int f(\tau)^4 \dd \tau \simeq 0.561244...$ is a constant numerical prefactor, calculated using $f(\tau)$ from Eq.~\eqref{eq:fluct1}. This corresponds, experimentally, to the low density $\phi = 0.02$ case where data can all be rescaled onto a single curve using the same diffusion coefficient for all box sizes (Fig.~\ref{fig:fig1}-b). The rescaling indicates that for non-interacting systems, the same diffusion coefficient governs behaviour irrespective of the length and time scales considered.


At higher packing fractions, a simple rescaling of the data with the self-diffusion coefficient fails (Supp. Mat. Sec. 1.4), \textit{i.e.}, the time required for fluctuations to relax does not solely depend on how long it takes a \textit{single} particle to diffuse over a particular length scale. Instead, for dense systems, the relaxation of number 
fluctuations depends on the movement of multiple interacting particles, hinting that it is 
governed by 
collective phenomena. 
 
To relate $T$ to a diffusive phenomenon, we rearrange the expression for $T(L)$ to define a diffusion coefficient, that is naturally dependent on the box size, as 
\begin{equation}
    D(\Lbox) = \alpha_T \frac{\Lbox^2}{4 T(\Lbox)}.
    \label{eq:Dbox}
\end{equation} 
$D(L)$
can be understood as the analogy of $D(k)$ in spectroscopy measurements where $k$ is the wavenumber~\cite{cerbino2008differential,berne2000dynamic}.
We can calculate this diffusion coefficient directly from our experimental or simulation data, shown rescaled by $\Dhydro$ in Fig.~\ref{fig:Dstar}. 

The diffusion coefficient at a specific scale $D(\Lbox)$ exhibits 4 regimes that we first explore via theory (Supp. Mat. Sec. 2.4). %

\begin{itemize}
    \item For small $\Lbox$, $D(\Lbox)$ converges to 
the self-diffusion coefficient $D(\Lbox \ll \sigma) = \Dbulk$ and so rescaled curves $D(\Lbox)/\Dbulk$ should tend to 1. Note that $\Dbulk$ can be replaced by $\Dhydro$ to include hydrodynamics \textit{a posteriori} in the theory. 
\item  For intermediate box sizes, $\sigma \leq \Lbox \leq \xi$, we find relaxation times are independent of the box size, as $T(\Lbox) = \beta_T \sigma^2/4 \Dbulk$, such that $D(L)/\Dbulk = \alpha_T L^2/ \beta_T \sigma^2$, 
where the numerical prefactor $\beta_T = (4-\pi)/16 \simeq 0.054$ (black dashed line in Fig.~\ref{fig:Dstar}a). This corresponds to the hyperuniform-like regime where fluctuations only need to relax over a length scale equal to the particle diameter. In this regime, for high packing fractions ($\phi = 0.66$, purple), we observe oscillations in $D(L)$, likely resulting from short-range structural order. 
\item For larger lengthscales, analytically $D(L)$ reaches a maximum, due to an interplay between collective dynamics speeding up the relaxation of fluctuations and fluctuations increasing in amplitude as they reach the entropic regime. 
\item Finally, in the limit of large boxes, $D(\Lbox \gg \xi) = \Dbulk/S(k=0) \equiv \Dstar$ reaches a second plateau, whose expression corresponds to the collective diffusion coefficient for systems with purely steric interactions  ~\cite{gomer1990diffusion,hess1983generalized,lahtinen2001diffusion,bleibel20153d} (Supp. Mat. Sec. 2.6). 
Numerical simulations without hydrodynamics (stars in Fig.~\ref{fig:Dstar}) agree with the theory, but plateau to a slightly higher value for intermediate density values. This discrepancy could originate from large density fluctuations which are not accounted for in our theory and in the derivation of $\Dstar = \Dbulk/S(k=0)$~\cite{bleibel20153d,dhont1996introduction}.
\end{itemize}

\noindent Therefore, $D(\Lbox) \sim \Lbox^2/T$ characterizes the self-diffusion coefficient on small boxes $L \ll \sigma$, while on large boxes, $L \gg \sigma$, it probes collective diffusion.

We now compare this behaviour with that of experiments and simulations that include hydrodynamic interactions. For small box sizes, rescaled values of $D(L)/\Dhydro$ computed from Eq.~\eqref{eq:time} also tend to 1, indicating that self-diffusion alone governs dynamics at this scale for all times. 
As $\Lbox$ increases, $D(L)$ also increases as with the theory. 
At high packing fractions ($\phi = 0.66$) and for $L \sim \sigma$, $D(L)$ exhibits slightly dampened oscillations compared to the theory, showing that even for boxes with just 1-3 particles, hydrodynamics have a non-trivial effect. 
Finally, for $L \gg \sigma$, $D(\Lbox)$ for systems with hydrodynamic interactions is increased by a factor of $2 - 10$, compared to the theory that only accounts for sterics. Thus, hydrodynamic interactions speed up diffusion and relaxation times significantly in this collective regime. This is in stark contrast with self-diffusion in dense systems which is reduced by hydrodynamic interactions with respect to the infinite dilution case.


Collective diffusion coefficients in colloidal systems have most commonly been probed in scattering experiments by measurement of the wavevector-dependent diffusion coefficient, $D(k)$, from the dynamic structure factor. For bulk 3D systems, it is generally accepted that hydrodynamic interactions reduce the value of both self \textit{and} collective diffusion coefficients~\cite{qiu1990hydrodynamic, segre1995experimental, bleibel20153d}. Yet for quasi-2D geometries, previous theoretical and experimental works have suggested that long-range correlations between particles can enhance collective motion dramatically, resulting in a divergence of $D(k)$ at small wavevectors~\cite{lin1995experimental,lin2014divergence,bleibel2014hydrodynamic,bleibel20153d,pelaez2018hydrodynamic,falck2004influence}. Interestingly, while we observe a clear enhancement of $D(L)$ arising from hydrodynamic interactions between particles, our data shows no sign of divergence over the range of box sizes accessible to us. This is likely a consequence of the different geometry of our system (particles at a single wall as opposed to a fluid-fluid interface or between two closely-spaced walls, as in the aforementioned works), as this is known to significantly influence the range over which hydrodynamic interactions decay~\cite{hashemi2022computing,Liron1976,FaxenThesis}. The relative sparsity of experimental studies on collective diffusion to date means that a systematic comparison between different geometries is, however, lacking. 
The timescale integral in the Countoscope thus offers a new way to shed light on the origins of this hydrodynamic enhancement of collective dynamics. 

\section*{Discussion}

In both experiments and simulations, the importance of \textit{static} number fluctuations for characterizing the thermodynamic properties of dense states has long been recognized~\cite{mahdisoltani2022nonequilibrium,zhang2010collective,peruani2012collective,liu2021density,toner2005hydrodynamics,fily2012athermal,dey2012spatial,fadda2023interplay,villamaina2014thinking,lebowitz1983charge,kim2005screening,chandler2005interfaces,kirkwood1951statistical,kusalik1987thermodynamic,schnell2011calculating,dawass2019kirkwood,cheng_computing_2022,torquato2018hyperuniform,hexner2015hyperuniformity}. \textit{Dynamic} number fluctuations are a natural extension to this, representing a similarly promising tool that has yet to be explored~\cite{cheng2020computing,minh2023ionic}. Here, by exploiting dynamic fluctuations, we introduce a robust, broadly applicable framework for probing diffusive properties in a wide range of systems. Given that our experimental geometry is commonly used, our findings for hard spheres represent a crucial first step towards understanding more complex particulate suspensions. Moving forward, we trust our analytical approach can be extended to 3D
, to solids or crystals, and to diverse interparticle interactions via structure factor models~\cite{minh2022frequency,minh2023ionic,te_vrugt_classical_2020,hansen2013theory}. Moreover, our counting methodology does not, in principle, require equilibrium conditions, hinting that out-of-equilibrium systems are also amenable 
with this framework. Generally, counting is an extremely sensitive tool as any dynamical feature, such as drift, long-time diffusion, or motion in the 3rd dimension, should be reflected in the fluctuating counts. Going forwards, this implies that many more dynamical properties can be extracted from fluctuating counts. 

The Countoscope complements existing techniques by enabling the straightforward analysis of dense and heterogeneous systems or arbitrary mixtures, as the identification of different species occurs in the position-finding step. In contrast, spectroscopy techniques are often limited to mixtures of only two types of particle, with one concentrated and one dilute component, due to the difficulty of establishing models to connect diffraction patterns and particle properties~\cite{hassan2015making,wilson2011differential,rose2020particle}. Microscopy techniques and simulations involving trajectories can not easily access dynamics in heterogeneous states since particles continuously enter and exit regions of interest. The Countoscope, however, can harvest these fluctuating numbers. Collective dynamics are hard to resolve for the same reason, as they require simultaneously following many particles for extended periods of time. As such, quantifying collective dynamics from experimental data remains an indisputable challenge, despite growing interest in collective phenomena in soft and biological systems~\cite{hallatschek2023proliferating,allen2018bacterial,marchetti2013hydrodynamics,alert2022active,be2019statistical}. The Countoscope naturally lends itself to probing collective properties and thus
offers an important opportunity to determine accurately the relaxation of fluctuations in the collective
regime, in a broad variety of systems. 

\section*{Appendix}

\subsection*{Appendix A: Experimental details}
Experiments were performed with highly monodisperse carboxylate-functionalized melamine formaldehyde particles (Microparticles GmbH) with diameter $\sigma = 2.8~\mu$m. The particles are dispersed in a 20/80 v/v$\%$ water ethanol mixture and suspensions are loaded into quartz glass flow cells. The high mass density of the particles ($\rho=1510$~kgm$^{-3}$) results in their rapid sedimentation to the base of the cell to form a monolayer geometry with packing fraction $\phi=N\pi \sigma^2/4A$, where $N$ is the number of particles and $A$ the area imaged. As the gravitational height of the particles is very small (approximately $2\%$ of the particle diameter) out-of-plane fluctuations of the particles are negligible. The sample cell is macroscopically large in the plane of the sample, but importantly also has a height of at least 70 times the particle diameter. As such, from a hydrodynamic perspective, particles only interact significantly with the base of the cell.  

Samples are imaged at a rate of 2~fps using a custom-built inverted microscope with a field of view $217\times 174~\mu$m$^2$. Measurements of the mean square change in particle number (as in Fig.~\ref{fig:fig2}) were initially resolved from datasets recorded over 20 minutes. To accurately compute the time integrals in Fig.~\ref{fig:Dstar}, however, data was recorded for approximately 20 hours. Before computing the integral, datasets were checked for any systematic drift in particle displacement over the timescale of the measurement. This was achieved by verifying that the mean square displacement of particle position scaled linearly in time over the course of the measurement. 

\subsection*{Appendix B: Numerical simulations}
All numerical simulations in this work were performed using the lubrication corrected Brownian dynamics method described by some of us in the context of dense suspensions of active particles above a bottom wall \cite{sprinkle2020driven,MonolayerSedimentation}. In particular, hydrodynamic interactions were implemented using the fast `spectralRPY' method for doubly periodic geometries\cite{hashemi2022computing}, which allowed for very large system sizes to be considered here ($\mathcal{O}( 10^4)$ particles). In simulations where hydrodynamic interactions were neglected, the mobility matrix, $\overline{\Mob}$, was taken to be a multiple of the identity specified by the self diffusion coefficient $D_0$, while the same `firm sphere' steric forces\cite{sprinkle2020driven} from the hydrodynamic simulations were used to prevent overlaps (particle-particle and particle-wall).  

The only input parameters to the simulations 
were a timestep $\Delta t$, a periodic domain size $L_p$, an initial configuration of particles, and the physical properties of the fluid - taken from experimental measurements as viscosity $\eta = 1.4 \times 10^{-3}$~Pa.s, temperature $T=22^{\circ}$C, particle diameter $\sigma = 2.79~\mu$m, and particle buoyant force $(\pi/6) \, \sigma^3 \, \Delta \rho \, g = 5.92 \times 10^{-14}$~N. To ensure that these parameter values produced simulated particle dynamics consistent with experiments, we compared the trajectory-based MSDs and observed excellent agreement between simulated and experimental measurements (see Supp. Mat. Sec. 1.1). 

Initial configurations for the suspensions considered in this work were sampled using a Markov chain Monte Carlo (MCMC) method, where the number of particles $N_p$ and the in-plane packing density $\phi=N_p \pi \sigma^2 / 4 L_p^2$ determined the periodic system size $L_p$. The value of $N_p$ was the primary control parameter 
of simulation efficiency. Larger values of $N_p$ yield larger domain sizes and better statistics over the total simulated time - at the expense of longer run times and, thus, shorter simulated trajectories. These efficiency considerations were used to find a suitable system size on a case-by-case basis, with $L_p$ ranging from $288$ to $3200~\mu$m. 

The timescale for steric interactions between particles controls the timestep size used in our simulations. To resolve these dynamics, we chose $\Delta t$ empirically for each value of $\phi$ so that temporal accuracy remained smaller than statistical error. In simulations including hydrodynamics, we found $\Delta t \sim 0.125 - 0.5$~s. In simulations without hydrodynamics, which are known to attenuate the large steric exclusion forces, we found that $\Delta t$ needed to be much smaller and, typically $\Delta t \sim 10^{-4} - 10^{-2}$~s. To partially ameliorate simulation efficiency in spite of this timestep reduction, 
we employ the time integration scheme introduced by Leimkuhler and Matthews\cite{LeimkuhlerAndMatthews}, which capitalizes on additive Brownian noise.


\subsection*{Appendix C: Particle number fluctuations analysis}

\subsubsection*{Calculating $N(t)$ from numerical and experimental data}

Data analysis is performed on sequences of the 
unlabeled $x$ and $y$ coordinates of particle centres at time (frame number) $t$ obtained from segmented experimental images and simulations. The total number of particles may change from frame to frame. The field of view, assumed to be fixed in time, is divided into $M$ tiles based on the specified observation box length $L$, and a specified separation width $\delta L$ between boxes. We used this separation to ensure that the statistics measured from each box are uncorrelated, which is obtained simply by taking $\delta L \simeq \sigma$. 
We count the particles whose centres fall within each observation box, generating a time series of particle counts $N_i(t)$ for box indices $i = 1,\ldots,M$. 

Statistics such as $\brak{\DN^2(t)}$ are estimated by first averaging over time origins $t_0$ then using arithmetic means over all of the boxes $\brak{\brak{\DN_i^2(t)}_{t_0}}_i$ and confidence intervals are estimated using 2 standard deviations over all of the boxes. The average particle number $\brak{N}$ and variance $\brak{N^2} - \brak{N}^2$ are estimated using the sample mean and unbiased sample variance, computed using all of boxes and all of the frames. For a few data points ($\phi = 0.66$, largest boxes), the plateau of $\brak{\DN^2(t)}$ does not agree with the sample variance due to large statistical errors, and we use the plateau of $\brak{\DN^2(t)}$ as a proxy for the variance. 

Given the above estimates for $\brak{\DN^2(t)}$ and $\brak{N^2} - \brak{N}^2$, the timescale integral in equation \eqref{eq:time} is calculated using trapezoidal summation. 
When the observation box size is very small, the integrand in Eq.~\eqref{eq:time} decays very quickly, and much of its support may fall before the first frame. 
When the observation box is very large, the decay is slow and we may not have enough data to calculate the integral accurately. Both of these issues can be addressed by fitting our available data at short and long times to theory-informed functional forms and calculating the missing contributions to the timescale integral. This procedure is described in detail in Supp. Mat. Sec. 1.5 which also includes a discussion of how the errorbars were calculated for Fig.~\ref{fig:Dstar}. 


\subsubsection*{Early time fits to obtain short time self-diffusion coefficients}

Fitting to Eq.~\eqref{eq:fit} can be used to estimate the value of $D_0^{\text{hydro}}$. For more accurate predictions, especially on small box sizes, one can instead expand Eq.~\eqref{eq:fluctk} at short times to second order in time and obtain
\begin{equation}
 \langle  \Delta N^2(t) \rangle = \frac{8}{\sqrt{\pi}} \sqrt{\frac{D_0^{\text{hydro}} t}{L^2}} -  \frac{8}{\pi} \frac{D_0^{\text{hydro}} t}{L^2}.
\end{equation}
Then, an estimate of $D_0^{\text{hydro}}$ can be obtained by inverting the above equation, which yields Eq.~\eqref{eq:fit2ndorder}. To fit to these equations, the average number of particles $\Nmean$ in a box is required. As correlations in number fluctuations decay slowly in time (as $1/t$), a simple estimate of $\Nmean$ averaging over all boxes and all times can be quite inaccurate. In general, we find better results by computing the total number of particles on each frame, averaged over time, $N_{\rm tot}$ and the mean number of particles per box as $\Nmean = N_{\rm tot} L^2/L_{\rm field}^2$. 

We obtain a value of $\Dhydro$ by averaging over boxes logarithmically-spaced between the smallest possible size of a pixel ($0.25~\mathrm{\mu m}$) and $32~\mathrm{\mu m}$; a box size which still allows one to fit 5$\times$5  boxes in the field of view. Estimations reported in Table~\ref{tab:fits} initially consider 60 box sizes in this range. From this set, we select a subset of box sizes for which we expect the estimator to be sufficiently accurate, according to rules outlined in SI Sec. 1.7.

\subsection*{Appendix D: Some analytical derivations}

\subsubsection*{Stochastic density field theory}

Detailed steps to obtain the analytic expressions above are reported in the Supplementary Material. Here we briefly recapitulate the main steps of the theory. We consider an ensemble of particles with standard repulsive interactions given by the interparticle potential $\VV(r)$ where $r$ is the distance between particles. 
 Let $\rho(\bm{r},t)$, where $\bm{r} = (x,y)$, be the particle number density field, so that the instantaneous number of particles the observation square box of side $L$ is
\begin{equation}
    N(t) = \int_{0}^L\int_{0}^L \rho(x,y,t) \dd x \dd y .
\end{equation}
While there are diverse strategies to calculate the statistical properties of $N$~\cite{smoluchowski1916studien,aebersold1993density,bingham1997estimating,goldenshluger2021smoluchowski,schnell2011calculating,marbach2021intrinsic,te_vrugt_classical_2020}, stochastic Density Functional Theory (sDFT)~\cite{dean1996langevin,kawasaki1994stochastic} stands out here for its simplicity. sDFT directly describes the fluctuations on the continuous field $\rho(\bm{r},t)$ due to individual particle diffusion and has been successfully applied in diverse systems to extract kinetic properties~\cite{demery2016conductivity,jardat2022diffusion,mahdisoltani2021transient,minh2023ionic,bernard2023analytical}. In addition, compared to the original techniques, \textit{e.g.} used by Smoluchowski~\cite{smoluchowski1916studien}, which were based on tracking individual particle jumps~\cite{marbach2021intrinsic}, sDFT can easily account for particle interactions~\cite{minh2023ionic}.

Our starting point is the Dean-Kawasaki equation~\cite{dean1996langevin,kawasaki1994stochastic} for the particle density $\rho(\bm{r},t)$
\begin{equation}
\begin{split}
    \frac{\partial \rho(\bm{r},t)}{\partial t}& = D_0 \bm{\nabla}^2 \rho \\
    &+ D_0 \bm{\nabla} \cdot \left( \rho(\bm{r},t) \bm{\nabla} \int \dd^2 r' \rho(\bm{r}',t)  \frac{\VV(|\bm{r}- \bm{r'}|)}{\kT} \right) \\
    &+ \bm{\nabla} \cdot (\sqrt{2D_0 \rho(\bm{r},t)}
 \bm{\zeta}(x,y,t) ) 
\end{split}
\label{eq:DeanKawazaki}
    \end{equation}
where $\bm{\zeta}$ is a 2-component gaussian noise, such that $  \langle \zeta_i(x,y,t) \rangle = 0$ and $\langle \zeta_i(x',y',t') \zeta_j(x,y,t) \rangle = \delta_{ij} \delta(x-x') \delta(y-y') \delta(t-t')$.

By linearizing Eq.~\eqref{eq:DeanKawazaki} around the mean particle density, and using standard approximations for interparticle interactions such as the random phase approximation~\cite{hansen2013theory}, we can obtain analytic expressions for particle number statistics in time.

\subsubsection*{Structure factor of hard spheres in 2D}

In Eq.~\eqref{eq:fluctk} and onwards we use an analytic expression for the structure factor of hard spheres in 2D, which is based on density field theory and is in remarkable agreement with our 2D-sedimented colloidal experiments, as was verified in a previous work~\cite{thorneywork2018structure}. We report it here for consistency:
\begin{equation}
    S(k) = \frac{1}{1 - \rho c^{(2)}(k)}
\end{equation}
where $\rho$ is the mean particle density and 
\begin{equation}
\begin{split}
        c^{(2)}(k) = &\frac{\pi}{6(1-\phi)^3 k^2} \bigg[ - \frac{5}{4} (1-\phi)^2 k^2 \sigma^2 J_0(k\sigma/2)^2 \\
        &+ \bigg( 4 \left( \,( \phi - 20) \phi + 7 \right) \\
        & + \frac{5}{4}(1 - \phi)^2 k^2 \sigma^2 \bigg) J_1(k\sigma/2)^2 \\
        &+ 2(\phi - 13)(1 - \phi) k \sigma J_1(k\sigma/2) J_0 (k\sigma/2) \bigg]
\label{eq:Sk}
\end{split}
\end{equation}
where $J_i(x)$ are Bessel functions of the first kind. The limit of vanishing wavenumber in Eq.~\eqref{eq:Sk} can be taken analytically
\begin{equation}
    S(k = 0) = \frac{(1- \phi)^3}{1 + \phi}
    \label{eq:sk0}
\end{equation}
and is consistent with the result from the scaled particle theory equation of state.


\section*{Data availability}

All data needed to evaluate the conclusions in the paper are present in the paper and/or the Supplementary Materials. All other data are available upon reasonable request to the authors.

\section*{Code availability}

Upon publication the code to analyze particle number fluctuations will be made available on Github. Simulation codes to generate particle trajectories are available at \url{https://github.com/stochasticHydroTools/RigidMultiblobsWall/tree/master/Lubrication}

\section*{Acknowledgements}

We wish to acknowledge fruitful discussions with  Ludovic Berthier, Adam Carter, Aleksandar Donev, Roel Dullens, Daan Frenkel, Simon Gravelle, Jean-Pierre Hansen, Pierre Illien, Marie Jardat, Pierre Levitz, Th\'{e} Hoang Ngoc Minh, Ignacio Pagonabarraga and Benjamin Rotenberg. S.M. is especially grateful to Federico Paratore for a very helpful conversation on an exploratory cruise in California. B.S. and S.M. thank Johanna McCombs for her early exploratory work on this subject and Sarah A. Hughes for her invaluable help with figures. S.M. received funding from the European Union’s Horizon 2020 research and innovation program under the Marie Skłodowska-Curie grant agreement 839225, MolecularControl. A.L.T. acknowledges funding from a Royal Society University Research Fellowship (URF\textbackslash R1\textbackslash211033). E.K.R.M. and A.L.T acknowledge funding from EPSRC (EP/X02492X/1).  B.S. acknowledges funding from the National Science Foundation award DMS-2052515. 

\section*{Author contributions}

The authors confirm their contribution to the paper as follows: study conception and design: S.M., B.S., A.L.T., data collection: E.K.R.M., B.S., A.L.T.; modeling: S.M.; data analysis: All; data interpretation: S.M., B.S., A.L.T.; visualization: B.S.; draft manuscript
preparation: S.M., A.L.T.; review and editing: S.M., B.S., A.L.T.

\section*{Competing interests}

The authors declare no competing interests.

%



\end{document}